# Laboratory biases hinder Eco-Evo-Devo integration: hints from the microworld

Natsuko Rivera-Yoshida[1,2,3], Alejandra Hernández-Terán[1,2], Ana E. Escalante[1], Mariana Benítez[1,3]*

[1]Laboratorio Nacional de Ciencias de la Sostenibilidad (LANCIS), Instituto de Ecología, Universidad Nacional Autónoma de México, Mexico City, Mexico

[2]Programa de Doctorado en Ciencias Biomédicas, Universidad Nacional Autónoma de México, Mexico

[3]Centro de Ciencias de la Complejidad, Universidad Nacional Autónoma de México, Mexico City, Mexico

Corresponding author: mbenitez@iecologia.unam.mx

**Abstract**

How specific environmental contexts contribute to the robustness and variation of developmental trajectories and evolutionary transitions is a central point in Eco-Evo-Devo. However, the articulation of ecological, evolutionary and developmental processes into integrative frameworks has been elusive, partly because standard experimental designs neglect or oversimplify ecologically meaningful contexts. Microbial models are useful to expose and discuss two possible sources of bias associated with gene-centered experimental designs: the use of laboratory strains and laboratory environmental conditions. We illustrate our point by showing how contrasting developmental phenotypes in *Myxococcus xanthus* depend on the joint variation of temperature and substrate stiffness. Microorganismal development can provide key information for better understanding the role of environmental conditions in the evolution of developmental variation, and to overcome some of the limitations associated with current experimental approaches.

## Eco-evo-devo at the microscale

Understanding the multicausal origins of biological variation constitutes a longstanding question, and interest in the variation generated by developmental processes occurring in different environmental conditions, as well as its evolutionary significance, is not new (Schmalhausen, 1949; Gupta & Lewontin, 1982; Scheiner & Goodnight, 1984; Sarkar 2004). However, since genetic variation has been considered the major cause of phenotypic variation, the organismal interaction with varying environments has been often reduced to noise or to deviations from a norm, assuming a univocal genotype-phenotype relation (Lewontin, 2001; Robert, 2004; Sultan, 2017). As such, the goal of much research in developmental biology has been to describe the effects of genetic differences on phenotype.

To reveal the phenotypic effects of genetic change, the dominant experimental approach has included two key elements. First, experimental designs have relied on a limited set of model organisms, sometimes restricted to particular laboratory lines or strains of those species (Ashburner et al., 2000; Robert, 2004; Kaletta & Hengartner, 2006). Second, these studies have deliberately excluded realistic environmental variation, instead rearing organisms in controlled, constant conditions that may be both very different from and much more stable than those in natural environments (Gilbert and Bolker, 2001). Although these approaches have generated a wealth of valuable results, they have also limited biological understanding to the extent that (a) model organisms do not capture key aspects of biological diversity, and (b) laboratory conditions intentionally restrict potential effects of natural environments (Bolker, 1995; Gilbert, 2001; Minelli & Baedke, 2014; Gasch et al., 2016). Thus, some important questions about the developmental and evolutionary processes occurring within specific ecological contexts have remained unsolved, or even unaddressed. For instance, how do both plastic and robust processes arise during development under varying natural contexts? What are the mechanisms by which phenotypic plasticity itself shapes ecological interactions? How often and how strongly does plasticity contribute to evolutionary processes such as phenotypic innovation in natural populations?

Considering these questions, there has been a recent increase in attention to phenotypic variation beyond single genetic sources and the role of the environmental context of phenotype expression and evolution (Moczek et al. 2011; Bateson and Gluckman 2012; Levis & Pfennig, 2016). This has given rise to the “Eco-Evo-Devo” approach, which emphasizes the relevance of the reciprocal interactions between ecology, evolution and development and focuses on multicausal development occurring in the “real-world” (Gilbert, 2001; Sultan, 2003). Despite the great efforts and progress on the conceptual approaches of Eco-Evo-Devo, the integration of the three fields that conform it is not yet complete due, at least in part, to the very nature of traditional experimental designs.

Microorganisms have been invaluable laboratory models in biotechnology and genetic research because they have short generation times and small size, are relatively easy to manipulate genetically and control experimentally, and are resistant to long-term storage (Jessup et al., 2004; O’malley et al., 2015). However, they are largely missing from Eco-Evo-Devo efforts, which have focused mainly on plants and animals, perhaps mostly for practical reasons such as ease of evaluating phenotypic outcomes (Gilbert, 2015; Sultan, 2015). Microorganismal growth and development offer remarkable examples of the restrictions implied by studies based on model organisms and standard laboratory conditions, which tend to cancel, minimize or underestimate the causal role of environmental variation in development. In fact, microorganisms represent huge biological diversity in terms of their metabolic capabilities and the wide range of ecological contexts they habit (Johri et al., 2005). Nevertheless, only approximately 1 % of the microbial diversity has been cultured using these experimental design strategies, illustrating our limited understanding of the organism-environment interaction required to reproduce microbial species (Pham & Kim, 2012; Nai & Meyer, 2017; Pande & Kost, 2017).

Furthermore, in light of their ubiquity across environments and their diverse uni- and multi-cellular lifestyles, microorganisms can provide invaluable insights to further understanding organism-environment interactions and the processes generating the variation that enables evolution. Moreover, multicellular microbial groups can yield information about the organism-environment interactions during the evolution of multicellularity since they develop in a scale and environment similar to those in which multicellularity presumably emerged (Bonner, 2009; Arias Del Angel et al., 2017; Rivera-Yoshida et al., 2018). Focusing on microorganisms also leads to the study of environmental variables that are less evident or relevant at the macroscale, such as the mechanical properties of cell-to-cell and cell-to-medium interactions (Persat et al., 2015).

Overall, microbial models can help to unmask biases in experimental designs implemented mostly in plants and animals, contributing to a better integration and experimental planning within the Eco-Evo-Devo field. Here, we focus on microbial models to illustrate how gene-centered experimental designs harbor two possible sources of bias: 1) the use of laboratory strains, and 2) laboratory environmental conditions. In particular, we gather evidence from different microorganisms and use our own results from *Myxococcus xanthus* development under different environmental conditions to exemplify and comment on these biases. In our opinion, such biases should be explicitly considered when interpreting results and extrapolating them to natural contexts and, ideally, should be overcome in novel empirical approaches.

**Laboratory standard strains vs. natural populations**

Choosing a model organism is often limited to some well-established options. Among microorganisms, *Escherichia coli*, *Bacillus subtilis* and *Saccharomyces cerevisiae* are widely used models (Love & Travisano, 2013; Blount, 2015). The use of laboratory standard strains has undeniable practical advantages that, in turn, reinforce the use of particular strains and species. These advantages include, for instance: 1) existing important technological investment including complete genome sequencing, protein and metabolite quantification methods, and mutant construction; 2) pure genetic lines and robust phenotypes that have been domesticated to grow under simple and standardized laboratory conditions; 3) minimal variation, which leads to tractable, systematic and reproducible results, and thus reliable comparisons; 4) data and techniques that can be shared among research groups, since they correspond with standardized conditions, including strains; and 5) popularity and facilitation of acquiring funding (Ankeny & Leonelli, 2011; Leonelli & Ankeny, 2013; Gasch et al., 2016). These features are particularly useful in exploring genetic mechanisms, since they help control the influence of non-genetic factors.

Nonetheless, while using standard strains can be of great value in microbiology, molecular biology, and some evolutionary studies, it becomes a limitation for other scientific purposes, such as those related with the Eco-Evo-Devo framework. Since biological questions should match the model organism and experimental decisions, standard strains are not well-suited to questions about phenotypic plasticity and its mechanisms, since plasticity has been intentionally or indirectly suppressed through invariant environmental conditions in already relatively unplastic organisms (Travis, 2006; Love, 2010). Indeed, model organisms often exhibit rapid development and developmental canalization (expression of a specific developmental outcome regardless of minor variations in environmental conditions;

Waddington, 1942), a well-known phenomenon in animal models (Bolker, 1995; Gilbert, 2001).

“Domestication” is commonly used to refer to the adaptation of wild strains to new, human-created habitats. When laboratories are the new habitats, domestication occurs in long-term, stable cultures or during repeated passaging (Branda et al., 2001; Kuthan et al., 2003; Palková, 2004; Eydallin et al., 2014). For microorganisms, laboratory-domesticated strains express robust phenotypic traits and apparently decreased phenotypic plasticity compared to strains that have been manipulated in the short term (Eydallin et al., 2014). However, whether these traits are actually canalized or not remains to be explored as reaction norm experiments are just starting to become available for microbial systems (Rivera-Yoshida et al., 2019). Thus, while domesticated strains enable important scientific and technical advances, relevant variation possibly occurring naturally at ecological complex scenarios, and the causes behind it, could be encrypted in these strains (Branda et al., 2001; Kuthan et al., 2003; Palková, 2004; Eydallin et al., 2014; Steensels et al., 2019).

Laboratory domestication has been reported for several microbial species. Interestingly, several species have been observed to develop common phenotypic traits during domestication when exposed to similar experimental contexts (Table 1). For instance, in laboratory conditions, standard *E. coli*, *B. subtilis* and *S. cerevisiae* strains present a smooth biofilm phenotype compared with the rough one observed in wild type strains (Branda et al., 2001; Kuthan et al., 2003; Palková, 2004; Eydallin et al., 2014). Moreover, in these three cases the smooth phenotype is related to the loss of complexity in the extracellular matrix structure (Table 1). Also, pathogenic laboratory strains present lower virulence compared to the newly isolated strains (Heddleston, 1964; Barak et al., 2005; White & Surette, 2006; Sommerville et al., 2011). The phenotypic convergence shared between species that have undergone independent domestication processes is reminiscent of the well-known domestication syndrome observed in crops (Gepts & Papa, 2002; Burke et al., 2007).

It should also be considered that domestication and genetic modification processes also involve the unintended selection of non-target traits (e.g. Hernández-Terán et al., 2017). Laboratory strains for the study of microbial multicellularity are a clear example. Wild *Myxococcus xanthus* and *Bacillus subtilis* strains can develop complex resistance structures in response to adverse environmental conditions. For these multicellular structures to occur, social behavior is needed. However, experimental setups tend to select easily dispersed cells or colonies, and then grow them in unstructured liquid mediums, which is reported to be associated to a reduction in social behavior (Velicer et al., 1998; Aguilar et al., 2007). This domestication pathway may thus hinder collective organization and actually makes them suboptimal for the study of multicellular development (Aguilar et al., 2007).

While there can be some convergences or similarities, it is overall difficult to generalize about domestication processes and outcomes, since different dynamics underlie each specific case. For instance, populations or ecotypes of the same microbial species can be widespread in completely dissimilar environments, and can exhibit different domestication trajectories (Eydallin et al., 2014). Laboratory domestication processes, phenotypes, and metabolic changes depend on the ancestral strains, on physical and chemical properties of the culture medium (e.g. liquid medium versus hard agar plate), and how long they have been exposed to the culture medium (Eydallin et al., 2014). Finally, studying systems with standardized strains and environmental conditions has the objective of supporting reliable comparisons across different research groups. However, due to the sensitivity of microbial strains to small variations on experimental treatments and also due to their long laboratory life history, sublines of the same laboratory standard strain could present phenotypic and genetic differences (Bradley et al., 2016).

Comparisons between laboratory strains and generalization to wild strains should be done with caution, since domestication processes occurring in association to widespread experimental approaches could impose important biases. The rapid domestication of microorganisms to laboratory conditions highlights the importance of working with recently isolated wild strains, at least for some research questions. In some cases, we do not even know if phenotypes that are commonly observed in laboratory strains actually exist in nature and are ecologically and evolutionarily relevant. For example, while in laboratory conditions *Myxococcus xanthus* forms well-known multicellular structures called fruiting bodies, we are not sure about what fruiting bodies look like when they develop in their natural soil environments. Further studies considering repeated, well-documented and already ongoing lab-domestication processes could also contribute to a better understanding of organism-environment interactions, phenotypic variation and robustness in a wide phylogenetic context (Bradley et al., 2016).

**Laboratory settings vs. natural environments**

Laboratory strains are good proxies of their wild ancestors if comparisons of their phenotypes and genotypes are not biased due to their history of experimental manipulation. However, that can only occur if a) the phenotypic outcomes of these strains were invariant with respect to the environment, or b) if laboratory conditions mimic natural conditions. The latter is clearly an unrealistic assumption, because as soon as an organism is isolated in laboratory culture media, natural environmental variables are modified. Moreover, as explained above, experimental designs have focused not on re-creating natural environments, but on generating “controlled environments” in which selected variables (often genetic variables) can be modified within a constant background (Robert, 2004). In this approach, controlled environments are assumed to be “neutral”, but they are actually conformed by several biotic and abiotic components contributing to the organism-environment interaction, which in turn may give rise to particular phenotypes (Lewontin, 2001; Sultan, 2017).

Within this controlled-environment setting, development -and its plastic nature- cannot be fully understood since it represents only one specific set of a wide possible repertoire of environmental conditions. Furthermore, beyond the constant background, experimental settings where a single variable is selected for modification can also be misleading in at least three ways. First, selected variables may not be ecologically meaningful for the studied organisms and developmental moment. Second, these variables could be ecologically meaningful but tested in non-significative ranges. Third, selected variables and unconsidered ones could be dynamically interacting and modifying the whole developmental system (see, for instance, Box 1). Indeed, meaningful environmental features could be the result of additive effects and complex interactions among variables, but it has been usually considered convenient to test only a few “key” variables, mostly in independent experimental sets (Rivera-Yoshida, et al., 2019). This approach, commonly associated with reaction norm studies, leads to interpreting the environment as a sum of major variables and, consequently, to limited conclusions.

Microorganisms have been considered important experimental models partially due to their ease of manipulation (Jessup et al., 2004; Love & Travisano, 2013). However, the natural history of most species, even cultivable ones, is unknown. Thus, their successful growth in the laboratory is informative about their ability to adapt to laboratory conditions but not necessarily about their growth and development in ecologically meaningful ones. For instance, the design of culture media is specially focused on chemical components for nutrient supply, while other physical or ecological factors are often overlooked. Remarkably, choosing the correct media chemical properties is not an easy task and may itself uncover interesting environmental dependencies (Uphoff et al., 2001).

The uncultivability phenomenon can provide clues about meaningful variables and ranges of natural settings neglected in current experimental designs, for example, by contrasting experimental properties with natural ones. Here we identify some key experimental conditions that differ from natural contexts. Growth media are restricted to either solid agar plates or liquid cultures commonly kept at constant agitation, which in turn, is known to favor loss of social behavior after just a few generations (Velicer et al., 1998). For agar plates, stiffness is standardized by fixing the agar concentration, but phenotypic plasticity has been described for microbial development and growth at different substrate stiffness (Be'er et al., 2009; Guégan et al., 2014; Rivera-Yoshida et al., 2019; Box 1). Also, agar plates represent flat and unstructured surfaces, determining properties of microbial aggregates and films such as movement, size and surface tension (Persat et al., 2015; Rivera-Yoshida et al., 2018). Nutrient supply is constant at optimum concentration rates or at complete scarcity. Genome reduction -also known as genome streamlining- occurs in natural populations when interacting species are metabolically complementary but also in long-term laboratory conditions (Koskiniemi et al., 2012; Lee et al., 2012; Pande & Kost, 2017). Constant and high nutrient supply are often part of experimental conditions, which could partially explain genome streamlining. Environmental settings are also restricted to small and homogeneous areas, limited by instrument walls or growing space. In contrast, natural environments are heterogeneous at different scales, which could in turn drive or constrain collective phenomena. For instance, the multicellular phenotype of *M. xanthus* can be characterized at the single fruiting body structure, but also at the population level (Rivera-Yoshida et al., 2019; Box 1). The population spatial distribution is determined by the experimental conditions (i.e. size of the culture plate or flask), however, it is unknown whether phenotypic expression at the individual or population scales are expressed in large and complex media like soil.

Abiotic physicochemical factors such as temperature, humidity, pH, pressure, salinity, and oxygen concentration, among others, are usually kept constant. However, these are important parameters determining microbial interactions and metabolism (Pham & Kim, 2012). Some of our own unexpected results with *M. xanthus* show how variations in temperature and medium stiffness can strongly affect bacterial multicellular development (Box 1). Indeed, *M. xanthus* fruiting bodies exhibit contrasting phenotypes when developed at different substrate stiffness, so much so that at very low stiffness and standard temperature, no fruiting bodies are formed (Box Figure 1 (c)). At standard stiffness, temperature modification yields little phenotypic variation, which could easily lead to the immediate conclusion that temperature does not affect development in a significant way. However, temperature variation reveals drastically different phenotypes at non-standard stiffness conditions, widening the spectrum of phenotypic variation associated with stiffness change (Box Figure 1 (d)-(f)). The joint modification of these two factors renders a phenotypic diversity that could not have been expected from *M. xanthus* being grown at standard conditions, nor from reaction norm experiments considering a single environmental factor (Rivera-Yoshida et al., 2019).

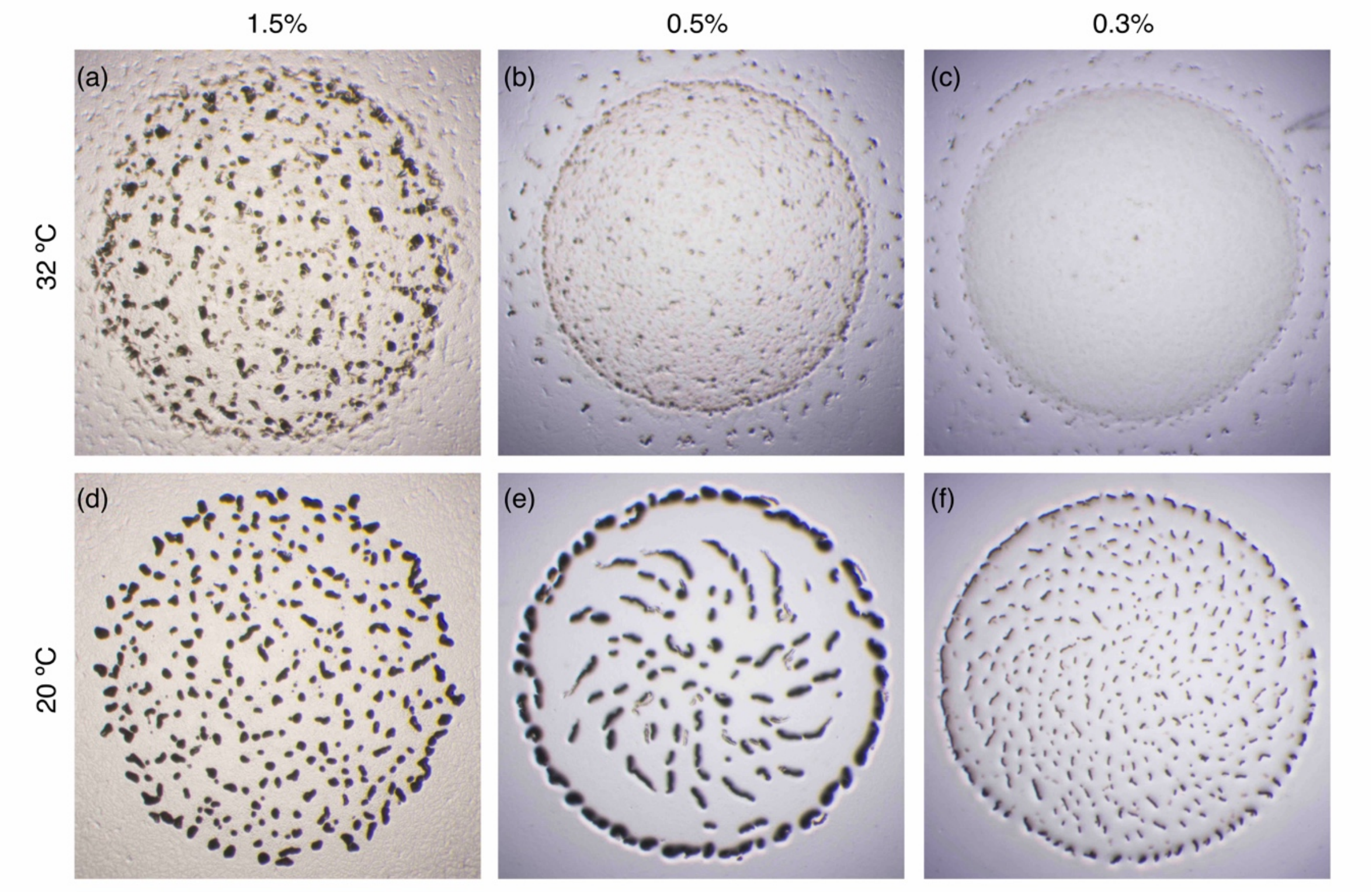

**Box 1**. Organism-environment interactions shaping multicellular development in *Myxococcus xanthus*. *M. xanthus* is a widespread soil bacterium with a multicellular developmental stage. It moves by gliding over semi-solid surfaces in the direction of the cell's long axis. In nutrient-rich substrates, cells are in a vegetative stage and swarm, expanding the colony outwards. When nutrients are depleted, they glide inwards, developing multicellular structures called fruiting bodies (FBs), where some cells eventually differentiate into resistant spores (Yang & Higgs, 2014). The standard protocol for multicellular development consists of depositing a drop of a liquid culture medium onto a nutrient depleted agar plate -commonly prepared with 1.5% agar-. After the drop dries, the plate is stored at 32 ºC avoiding light for about 96h until FBs have developed.

*M. xanthus* cells sense and respond to the structural and mechanical properties of the substrate over which they move. For example, they realign perpendicularly to mechanical compression applied to the agar mesh (Lemon et al., 2017; Fontes & Kaiser, 1999). Additionally, *M. xanthus* development has revealed two scales of phenotypic expression: the single FB scale and the population scale (the collection of FBs within a drop), both of which present phenotypic plasticity when substrate stiffness is modified (Rivera-Yoshida et al., 2019). The effect of other variables such as temperature, has not been widely or systematically tested. Furthermore, substrate stiffness is modified by varying the agar concentration. However, substrate mechanical properties might be the result of the interaction of more than this single variable. For instance, substrates with the same agar concentration but different temperatures, could differ in stiffness.

**Box Figure 1.** Phenotypic plasticity of *Myxococcus xanthus* multicellular structures. Micrographs of completely matured FBs populations developed over TPM agar plates. Black structures are FBs. The DZF1 standard laboratory strain was tested modifying temperature and agar concentration: (a) standard protocol condition: 32 ºC and 1.5% agar concentration. (b) 32 ºC, 0.5% (c) 32 ºC, 0.3% (d) 20 ºC, 1.5% (e) 20 ºC, 0.5% and (f) 20 ºC, 0.3%. Micrographs of each drop were taken at 370.8 pixels/mm using a LEICA m50 stereomicroscope with an ACHRO 0.63x objective lens and a Canon-EOS Rebel T3i camera. Apart from variation in temperature and agar percentage, *M. xanthus* were grown and developed as described in Yang and Higgs, 2014.

Besides the experimental substrate, other differences between experimental and natural settings can be associated with the management of biological material. Development or growth rates may be different among species, requiring longer or shorter periods to become visible to the experimenter. However, given the high nutrient supply, no more than a few days are given to cultures for their density to increase. Also, population densities are probably much higher than on natural substrates (Pande & Kost, 2017). In *M. xanthus*, multicellular development of fruiting bodies under conditions of nutrient scarcity is known to happen at high cell densities, around $1x10^4$ cells per fruiting body (Velicer et al., 1998), but actual cell density in natural substrates remains unknown. In the likely case that natural densities are much lower than experimental ones, what is known about developmental and quorum sensing mechanisms might be substantially different in natural populations. Additionally, axenic cultures are promoted in experimental designs so that species are intentionally isolated from interspecific interactions. Yet, the importance of dependence, predation and cooperation, among other interactions, for microbial growth and development are largely known (Jacobi et al., 1996; Pande & Kost, 2017). Finally, laboratory populations are mainly composed of clonal populations so that their genetic background lacks the heterogeneity observed in natural populations (Eydallin et al., 2014; Gasch et al., 2016).

The organism-environment interaction is a constantly changing bidirectional process, which also changes with spatiotemporal scale. In both natural and experimental settings, organisms contribute to the reconstruction of the inter-species niche (Miner et al., 2005; Ryan et al, 2016). For instance, bacterial extracellular matrix secretion is altered by the medium mechanical properties, which in turn are altered by the extracellular matrix secretion (Be'er et al., 2009; Fauvart et al., 2012; Trinschek et al., 2017; Rivera-Yoshida et al., 2018). Thus, dynamics associated with natural and experimental settings cannot be fully compared as they follow their own evolutionary tempos and paths. Complex ecological interactions are still far from laboratory proxies and efforts to improve protocols in the field or alternative experimental designs that consider environmental complexity are thus necessary.

**Final remarks**

The microbial world has provided new insights and approaches in the study of organism-environment interactions at both the evolutionary and ecological levels (Jessup et al., 2004; Love & Travisano, 2013; O'malley et al., 2015; Rivera-Yoshida et al., 2018). However, developmental mechanisms have been only partially understood since they have been studied through the establishment of experimental designs using domesticated strains and invariable conditions. This approach is only informative about a specific and simplified condition from the wide repertoire of environmental settings occurring in nature, in which phenotypic plasticity mechanisms may be obscured. Nevertheless, observations of microbial development highlight the importance of commonly overlooked, yet meaningful properties of the environment at the microscale. For instance, mechanical factors affecting living and nonliving matter play a key role determining substrate properties, which in turn, modify organisms' dynamics, such as spread, movement and development (Persat et al., 2015; Rivera-Yoshida et al., 2018). Microbes' plastic responses to other ecological factors such the presence of predators, interspecies interactions or environment fluctuation remain largely unknown.

The use of laboratory models and conditions like the ones described above respond, at least in part, to the pressure on science to be efficient in terms of time and costs, which in turn favors certain experimental setups and approaches, including standardized organisms and experimental conditions (Levins & Lewontin, 1985; Ankeny & Leonelli, 2011; Leonelli & Ankeny, 2013). Compelled by these pressures, microbial ecological and evolutionary processes are probably forced into tempos and conditions that do not match those of natural environments, leaving some open questions. For example, what are the ecologically relevant spatiotemporal scales and variables for microbial development? Is the strength and expression of phenotypic plasticity scale-specific? How plastic are interspecific interactions? How do different environmental variables interact with each other to affect microbial development?

Furthermore, the role of phenotypic plasticity as a driver of and constraint on evolutionary mechanisms, considered in hypotheses such as "plasticity-first", is probably underestimated since it cannot be easily tested in current experimental designs nor compared with phenomena occurring in natural populations (Levis & Pfennig, 2016). For instance, to the best of our knowledge, environmentally-triggered phenotypic novelties and complex interactions among environmental variables have not been explored in microbial systems, not even in paradigmatic long-term evolutionary studies. Additionally, due to their high mutation rate and short generation time, microbial groups could be suitable to the comparison of adaptations driven by plasticity versus mutation. Overall, further investigating microbial multicellular development and considering the practical biases underlying its current study can provide invaluable insights for the integration of Eco-Evo-Devo and understanding of major transitions in evolution.

**Table 1.** Microbial strains commonly used in laboratory conditions. * No information found.

| Species | Natural habitat | Laboratory strain phenotype | Wild strain phenotype | Research focus | Laboratory strain limitations | References |
|---|---|---|---|---|---|---|
| *Bacillus subtilis* | Plant roots<br><br>Soil<br><br>Animal intestinal tracts | Simple macroscopic architecture<br><br>Thin, fragile, smooth biofilm | Structurally complex<br><br>Thick and rough biofilm | Molecular mechanisms of colony morphogenesis<br><br>DNA mediated transformation | Lack of surfactin production: no spreading behavior<br><br>Inability to form resistant spore structures: multicellularity cannot be fully studied<br><br>Loss of genes or mutations | McLoon et al., 2011<br><br>Hong et al., 2009<br><br>Aguilar et al., 2007<br><br>Branda et al., 2001 |
| *Escherichia coli* | Soil<br><br>Water<br><br>Plant tissues<br><br>Animal gut | Smooth biofilm | Structurally complex<br><br>Rough phenotype | Pharmaceutical production<br><br>Genetic engineering<br><br>Biotechnology industry | Changes in biofilm structure: loss against predators<br><br>Changes in metabolic properties | DePas et al., 2014<br><br>van Elsas et al., 2011<br><br>Blount et al., 2008<br><br>Tao et al., 1999<br><br>Mikkola & Kurland, 1992 |
| *Bartonella henselae* | Animal blood and | * | Fimbriae presence | Medical research | Mutations and genomic | Arvand et al., 2006 |
| | endothelium | | Population genetic variation | | rearrangement during laboratory passaging<br><br>Decreased genetic variability<br><br>Loss of fimbriae<br><br>Lower virulence | Vhelo et al., 2002 |
| *Staphylococcus aureus* | Human skin, bones, blood and mucous membrane | * | High growth yield<br><br>High ROS production<br><br>High fitness | Medical research | Mutations and genomic rearrangement during lab passaging<br><br>Lower virulence<br><br>Alteration in cell density sensing<br><br>Alteration of surfactant production | Periasamy et al., 2012<br><br>Somerville et al., 2011 |
| *Pasteurella multocida* | Animal lungs, bloodstream and mucous membrane | * | Loss of capsule production | Medical research | Loss of capsule production: critical for its virulence and antibiotics resistance | Steen et al., 2010<br><br>Harper et al., 2006<br><br>Heddleston et al., 1964 |
| *Salmonella enterica* | Birds eggs<br><br>Plant tissues<br><br>Water<br><br>Animals intestinal tract | Smooth biofilm | Structurally complex biofilm<br><br>Dry and rough biofilm | Infectious disease studies | Diversification of genotypes<br><br>Altered biosynthesis of cellulose and polysaccharides: loss of spatial phenotype morphology<br><br>Resistant to desiccation | Davidson et al., 2008<br><br>White & Surette, 2006<br><br>Barak et al., 2005 |
| *Saccharomyces cerevisiae* | Oak bark and other trees and plants | Smooth colonies<br><br>Change in | Structurally complex biofilm | Genetic engineering<br><br>Biotechnolog | Loss of extracellular matrix | Šťovíček et al., 2014<br><br>Piccirillo & |
| | Human microbiota<br><br>Insects<br><br>Soil<br><br>Bread<br><br>Beer and wine | cells shape | Highly glycosylated extracellular matrix<br><br>Resistant to antioxidative stress<br><br>Phenotypic heterogeneity<br><br>Sporulates on a wider range of carbon sources | ical industry | Gene expression reprogramming | Honigberg, 2010<br><br>Diezmann & Dietrich, 2009<br><br>Palková et al., 2004<br><br>Kuthan et al., 2003 |
| *Myxococcus xanthus* | Soil | Smooth colonies | Large genetic heterogeneity<br><br>Large variation in phenotypic and developmental traits | Multicellularity<br><br>Cell differentiation<br><br>Cell motility | Loss of social behavior: multicellularity cannot be fully studied | Kraemer et al., 2010<br><br>Velicer et al., 1998 |

**Acknowledgements** Natsuko Rivera-Yoshida is a doctoral student in the Programa de Doctorado en Ciencias Biomédicas, Universidad Nacional Autónoma de México (UNAM) and received fellowship 580236 from CONACYT. The authors thank members of LANCIS, Sonia Sultan, Marcelo Navarro-Díaz, Alejandro V. Arzola, Alessio Franci and Juan A. Arias Del Angel for their support and valuable feedback.